# Establishing Magnetic Coupling in Spin-crossover-2D Hybrid Nanostructures via Interfacial Charge-transfer Interaction

Shatabda Bhattacharya[1, 3], Shubhadip Moulick[1#], Chinmoy Das[2#], Shiladitya Karmakar[1#], Hirokazu Tada[3], Tanusri Saha-Dasgupta[1], Pradip Chakraborty[2*], Atindra Nath Pal[1*]

[1]*Department of Condensed Matter Physics and Material Sciences, S. N. Bose National Centre for Basic Sciences, Salt Lake City, Kolkata - 700 106, India*

[2]*Department of Chemistry, Indian Institute of Technology Kharagpur, Kharagpur-721302, India*

[3]*Department of Materials Engineering Science, Graduate School of Engineering Science, Osaka University, Toyonaka, 560-8531, Japan*

*Corresponding Authors: atin@bose.res.in; pradipc@chem.iitkgp.ac.in

(# contributed equally to this work)

**Abstract:**

Despite a clear demonstration of bistability in spin-crossover (SCO) materials, the absence of long-range magnetic order and poor electrical conductivity limit their prospect in spintronic and nanoelectronic applications. Intending to create hybrid devices made of spin-crossover (SCO)-2D architecture, here, we report an easily processable Fe-based SCO nanostructures grown on 2D reduced graphene oxide (rGO). The heterostructure shows enhanced cooperativity due to formation of interfacial charge transfer induced inter-molecular interaction. The spin transition temperature is controlled by tuning the coverage area of SCO nanostructured networks over the 2D surfaces, thus manipulating hysteresis (*aka* memory) of the heterostructure. The enhanced magnetic coupling of the heterostructure leads to the spontaneous magnetization states with a large coercive field of ~ 3000 Oe. Additionally, the low conductivity of the pristine SCO nanostructures is addressed by encapsulating them on suitable 2D rGO template, enabling detection of magnetic bistable spin states during high-spin/low-spin conductance change. This adds spin functionality in conductance switching for realizing hybrid 2D spintronic devices. Ab-inito calculations, on the experimentally proposed nanostructures, corroborate the enhanced magnetic interaction in the proposed architecture facilitated by interfacial charge transfer and provide insights on the microscopic mechanism.

**Introduction:**

At the molecular level, control of magnetic bistable states has attracted much interest[1–5]. Fundamentally, the bistability can be related to binary 0 and 1, the elementary bits used for information storage[3,4]. Hence, molecule-based materials with bistability, for instance, spin-crossover (SCO) materials and molecular nanomagnets have futuristic applications in high-density information-storage, molecular switches and spintronics[6–8]. Among numerous SCO materials, by far, pseudo-octahedral Fe(II) complexes are the best candidates that display reversible spin-state switching between the low spin (LS) and high spin (HS) states arising from the rearrangement of electrons within d-orbitals, leading to the configuration with minimum (*i.e.*, S=0, diamagnetic) and maximum (*i.e.*, S=2, paramagnetic) spin-multiplicity, respectively. This happens under the influence of various physicochemical external stimuli such as temperature, pressure, light, electric field, chemical variation etc.[9–11] The magnetic interaction between SCO centers in such materials are generally weak or negligible due to weak van der Waals or hydrogen bonded inter-molecular interactions. It has been observed that the SCO molecules, either in solution or in metal-diluted lattices, show a gradual spin-crossover behaviour. In contrast, in concentrated solids, the lattice effect comes into play that gives rise to the cooperativity among the spin-crossover active centers, originating from the elastic interaction that at short-range occurs *via* a considerable change in molecular volume between the HS and LS states and also manifested by the geometry, orientation, and packing of the neighbouring spin-crossover molecules, while at long-range, is mediated *via* electron-phonon coupling that depends on the average number of HS molecules [9,12–17]. In recent literature [16], the magnetic interaction between the SCO centres has also been argued to play an important role. This cooperativity is fundamentally related to hysteresis and, thus, memory. The bottleneck behind the application of SCO molecular materials in electronic and spintronic devices is the absence of long-range magnetic ordering and the electrically insulating behaviour, which we address in our study.

Though, cooperativity can be tuned by engineering the coordination site of the metal centers or through modification of elastic interactions by changing the ligand-fields or by varying the surrounding environment or dimensions [18–23], the cooperativity that is needed for application is yet to be achieved. The magnetic exchange interaction can be enhanced or be introduced via substrate incorporation and the presence of an extraneous matrix can affect the long-range cooperativity and modulate the SCO behaviour[9,24,25]. However, while depositing molecules on surface, the role of substrate must be cautiously considered, as SCO property is susceptible to subtle changes in its local environment[26,27], due to which, the device miniaturization on 2D substrate becomes difficult[28,29]. Unlike molecular crystals with isolated units, 3D molecular network gated by conductive 2D substrate are better candidates for extending the magnetic exchange interaction through chemical bridges. With this technique, magnetism in spin-crossover can be sustained until ambient temperatures.

Here, we report easily processable, robust SCO-2D hybrid heterostructures that show long-range magnetic ordering through charge transfer induced exchange interactions. Fe(II)-Triazole based SCO nanostructures ([Fe(Htrz)$_2$(trz)]BF$_4$; Htrz: 1,2,4-triazole) and chemically synthesized reduced graphene oxide (rGO) have been chosen for this hybrid heterostructures. Due to the many functional groups on the surface of rGO, it is a promising candidate for anchoring SCO nanostructure networks epitaxially on rGO without any additional bridging agent. The strategy results an unrivalled technique for optimizing magnetic and electrical properties. After the formation of the heterostructure, interfacial charge transfer (CT) between SCO and rGO has been experimentally evidenced *via* X-ray photoelectron spectroscopy (XPS). The CT promotes a long-range magnetic exchange coupling among the Fe(II) metal centers and induces magnetization of rGO. The interfacial interaction between 2D conducting template and SCO NPs changes the high-spin (HS) and low-spin (LS) transition states of the hybrid heterostructures by more than 40K for tuning memory of SCO since thermal hysteresis could be fundamentally related to memory. Earlier, photoinduced magnetically ordered state in SCO systems have been achieved only at low temperatures (<130K) by LIESST effect (light induced excited spin state trapping) [11,30]. But the magnetization parameters like coercive field and magnetic moment generated due to LIESST effect could have been more promising for practical realization (e.g., 240 Oe at 20K)[11]. Other Fe(II) based photoinduced nanomagnets failed to display magnetic hysteresis related to binary units[30–32]. However, in contrast, in the present work, rGO mediated π-d exchange interactions with the Fe-centers result in thermally stable, long-range ordering in the high temperature range (~340K). Moreover, we have found that the thermal hysteresis width (ΔK) and the spin-transition temperatures ($T_{1/2}^{up}$, $T_{1/2}^{down}$) can also be controlled enormously by tuning the coverage area of the SCO nanostructure network over the rGO surface. While pure [Fe(Htrz)$_2$(trz)]BF$_4$ nanoparticle shows a thermal hysteresis width of 39K, in-case of SCO-rGO hybrid heterostructures, a field-tunable variation in the hysteresis width of 10-60K has been optimized with the aim of using it as a wide bistable memory channel. In addition, the pristine [Fe(Htrz)$_2$(trz)]BF$_4$ has zero coercivity due to paramagnetic nature, whereas, in SCO-rGO heterostructure, we have observed a giant coercive field of 2800 Oe with long-range magnetic ordering, thanks to the disordered 2D matrix of rGO for generating such high surface-pinned coercive field. By changing the coverage area of Fe-Trz on rGO, the magnetic hysteresis loop changes from the ferromagnetic (cooperative) to antiferromagnetic (anti-cooperative) in nature. Apart from the evolution of spontaneous magnetization in the heterostructure, the spin state dependent electrical conductivity in Fe-Trz SCO is also primarily influenced due to high conducting nature of the 2D rGO surface. The typical insulating character (conductance ~ $10^{-11}$ S) of the pristine Fe-Trz SCO nanoparticles has been greatly improved after embedding them on rGO surface and the conductance is enhanced by six orders of magnitude (~$10^{-5}$ S). Since the two spin states have different conductance, implementing molecular spintronics switches can be facilitated by using these SCO-2D hybrid nanostructured materials.

Our experimental findings are further supported by the ab-initio density functional theory (DFT) calculations. DFT calculations identify the potential charge transfer pathways between SCO nanostructures and rGO with finite charge accumulation at the interface, which promotes effective inter-molecular interaction between SCO centres, facilitated by SCO and rGO interaction. Magnetic exchanges, calculated by mapping the DFT total energy of different spin configuration into a spin Hamiltonian, exhibit robust intra-chain super exchanges in the hybrid architecture.

**Experimental:**

**Synthesis of rGO/[Fe(Htrz)$_2$(trz)](BF$_4$) nanocomposite**

[Fe(Htrz)$_2$(trz)]BF$_4$ nanoparticle network was synthesized by the adapted reverse micelle technique previously used by Coronado et al.[24] and Roubeau et al.[25]. First, two separate microemulsions were prepared at room temperature. (i) An aqueous solution of Fe(BF$_4$)$_2$· 6H$_2$O (0.3 mL, 1 M) was prepared by dissolving 101.3 mg Fe(BF$_4$)$_2$.6H$_2$O in 0.3 mL deionized water, and a pinch of ascorbic acid was added to prevent Fe(II) oxidation. Subsequently, it was added to the previously prepared 10 mL of n-octane solution of sodium dioctyl sulfosuccinate (NaAOT) (1.35 g) and stirred for 30 min to produce a stable microemulsion. 0.402 g behenic acid was added as a co-surfactant and stirred for another 1 hr. (ii) For the second microemulsion, an aqueous solution of 1,2,4-triazole (Htrz) (0.3 mL, 3 M) ligand was prepared by dissolving 62.16 mg Htrz in 0.3 mL deionized water and the solution was subsequently added to another NaAOT (1.35g) and octane solution (10 mL) with constant stirring for 30 mins to get a stable microemulsion.

The two microemulsions were filtered, and the second microemulsion was added to the first one and stirred constantly. With time, the [Fe(Htrz)$_2$(trz)]BF$_4$ nanoparticles (NP) were formed through the micellar exchange, and the solution turned into a clear deep pink suspension. The reaction was quenched after 30 min by adding 50 mL acetone, followed by centrifugation and washing three times with ethanol to remove the excess surfactant. The washed nanoparticles are purple and finally dispersed in ethanol before attaching to reduced graphene oxide (rGO) surface. Before attaching the NPs to the rGO surface, we observed that the reversible color change occurs during the thermal spin transition using a simple heating/cooling method (see **Figure S1**).

Moreover, the as-synthesized nanoparticles' mean size and size distribution were measured using Dynamic Light Scattering (DLS). The resultant average size and distribution are 34.87 nm and 30-40 nm, respectively (see Figure S2). Additionally, the FTIR spectrum was obtained for the washed and aged nanoparticles, and the data was recorded in the PerkinElmer Spectrum Two spectrometer instrument (shown in **Figure S3**). The fingerprint of IR data is unambiguously matched with the previously reported one[25].

The effective composition (in wt. %) of the major constituent atoms (*i.e.*, C, N, Fe, and F) of the as-synthesized nanoparticles is confirmed by Energy-dispersive X-ray analysis (performed in FESEM-Jeol, Japan, JSM-J610F instrument, at an applied voltage of 15kV) and summarized in Table S1 and the corresponding spectrum is shown in **figure S4**, where H and B (not shown) atoms are below the threshold detection limit.

In the second step, 10mg of graphene oxide (GO) powder [purchased from *Graphenea*TM] was dispersed in 50 ml of ethanol with ultrasonication for two hours to exfoliate the layers. After this, a few drops of as-synthesized [Fe(Htrz)$_2$(trz)](BF$_4$) NPs were diluted in 10 ml of ethanol in a separate beaker. Next, the GO suspension was heated to 60°C under continuous stirring. To this, the [Fe(Htrz)$_2$(trz)](BF$_4$) nanoparticles were added slowly. The mixture was washed and filtered to remove any excess solvents. As GO has a large number of functional groups in its basal plane and the surface of the GO has a negative environment, the SCO nanoparticles can easily attach to the GO surface *via* electrostatic interactions. Also, they can intercalate within the interlayer separation of GO. Next, the composite was further thermally reduced in the presence of ultra-high pure (99.9%) H$_2$ gas in a closed quartz 3-zone CVD furnace at a temperature of 200°C for four hours to complete the reduction process and formation of the SCO-rGO heterostructure. We prepared two batches of samples by varying the SCO NP's coverage area on the rGO surface. One is the high coverage (concentration) of SCO nanoparticles on rGO, termed HC, and another is the very dilute concentration of SCO nanoparticles on rGO, termed LC. Other characterization techniques revealed that uniform 2D assemblies of SCO NPs have formed over the rGO surface.

**Theoretical:**

**DFT calculations and extraction of magnetic exchanges**

Ab initio calculation is carried out within the framework of density functional theory with exchange-correlation approximated through Perdew-Burke-Ernzerh formulation of generalized gradient approximation (PBE-GGA)[33] To take into account the strong correlation effect at Fe site, beyond PBE-GGA, we invoke the supplemented Hubbard U (DFT+U) approach. Calculations are carried out in plane wave basis set employing projector augmented-wave pseudo potentials, as implemented in the Vienna ab initio simulation package (VASP)[34]. Computational details are given in the supporting information.

To investigate the hybrid Fe-Trz-rGO nanostructure theoretically, we build up a model system to simulate the interfacial effect between corresponding SCO compound and rGO. Details about the simulation model and constructed nanostructures are given in the supporting information. For stability analysis of the hybrid composite, we compute the binding energy using formula: $E_{binding} = E_{model} - E_{chain} - E_{substrate}$, where $E_{model}$ denotes the total energy of the model system containing Fe-triazole chain embedded on rGO substrate. $E_{chain}$ and $E_{substrate}$ denote the total energy of Fe-triazole chain and rGO, respectively. According to the definition, the -ve value of $E_{binding}$ indicates favourable binding of the Fe-triazole chain on the substrate.

To extract the magnetic exchanges, total energy calculations are carried out considering different alignment of Fe spins. The energies are then mapped to Ising-like spin model Hamiltonian to extract the corresponding magnetic exchanges. Notably, only the intra-chain interactions have been extracted, assuming the inter-chain interactions to be negligible. While this assumption is true for high coverage of SCO chains, with intra-chain interactions arising from several chains dominating the inter-chain interaction, at low coverage the inter-chain interaction mediated by the rGO carriers are expected to show its presence, which has not been accounted for in the calculation.

**Results and Discussions:**

**Structural morphology by transmission electron microscopy (TEM):**

**Fig. 1(a)** illustrates the schematics of the hybrid SCO-rGO heterostructure to provide a simplified overview of the entire network and their connectivity. **Fig. 1(b)** shows the TEM micrograph of the pristine SCO nanoparticle network connected to each other. **Fig. 1(c)** depicts the overall morphology of the hybrid structure after the attachment of SCO nanoparticle-network over rGO surface. Thin rGO plane can be seen in the background over which the SCO nanoparticles are covered. From the transparency of rGO layer it can be concluded that it contains only few layers. In the high-resolution image of HC sample (**Fig. 1d**), SCO nanoparticles are densely distributed over rGO surface covering almost the entire area. The tunability in coverage area is changed in **Fig. 1(e)**, where the low concentration sample is shown. In LC, the dense overlapping is controlled and a uniform network is visible. No dense nanoparticle-network outside the rGO layer has been found in both the cases that verifies the controlling ability of the synthesis.

**Charge transfer and bonding state analysis by X-ray Photoelectron Spectroscopy (XPS):**

To detect the elemental oxidation states and functionalization process through a binding energy change, we have carried out XPS analysis of the SCO-rGO nanocomposite and compared it with pristine rGO by deconvoluting the high-resolution peaks. In-case of the SCO-rGO composite C 1s spectra (**Fig. 1f**), after deconvolution, it is found that in addition to usual carbon-containing groups, the formation of additional C-N bond (at around 286eV) is evidenced. It implies that the molecules bound to the surface of rGO through nitrogen ligand remain in the outermost part of the SCO structure [26]. While in the case of pristine rGO C1s spectra (**Fig. 1g**), no such peak position was found. Due to the attachment of SCO nanostructure, the most substantial C=C peak (due to $sp^2$ hybridized carbon atoms) is shifted to a higher binding energy side (*i.e.*, the peak appears at about 282.5 eV for rGO and shifts to 283.7 eV for SCO-rGO nanocomposite). The change in the binding energy is due to the charge transfer between rGO and the SCO nanostructures at the interface [27]. It is also pertinent to mention that after the attachment of SCO networks, the FWHM of C1s spectra is increased primarily due to attachment of the triazole-based compound to the carbon surface [27]. After the deconvolution of the Fe 2p states (**Fig. 1h**), two oxidation states of Fe have been found, which is typical for the Fe-containing Triazole compound. The peaks around 709.7 eV and 724.3 eV are due to $Fe^{2+}$ $^2p_{3/2}$ and $^2p_{1/2}$ levels, while the other peaks at around 714.4 eV and 728.9 eV are due to $Fe^{3+}$ levels[27,28]. Two small extra peaks are satellite spectra of the Fe 2p phase.

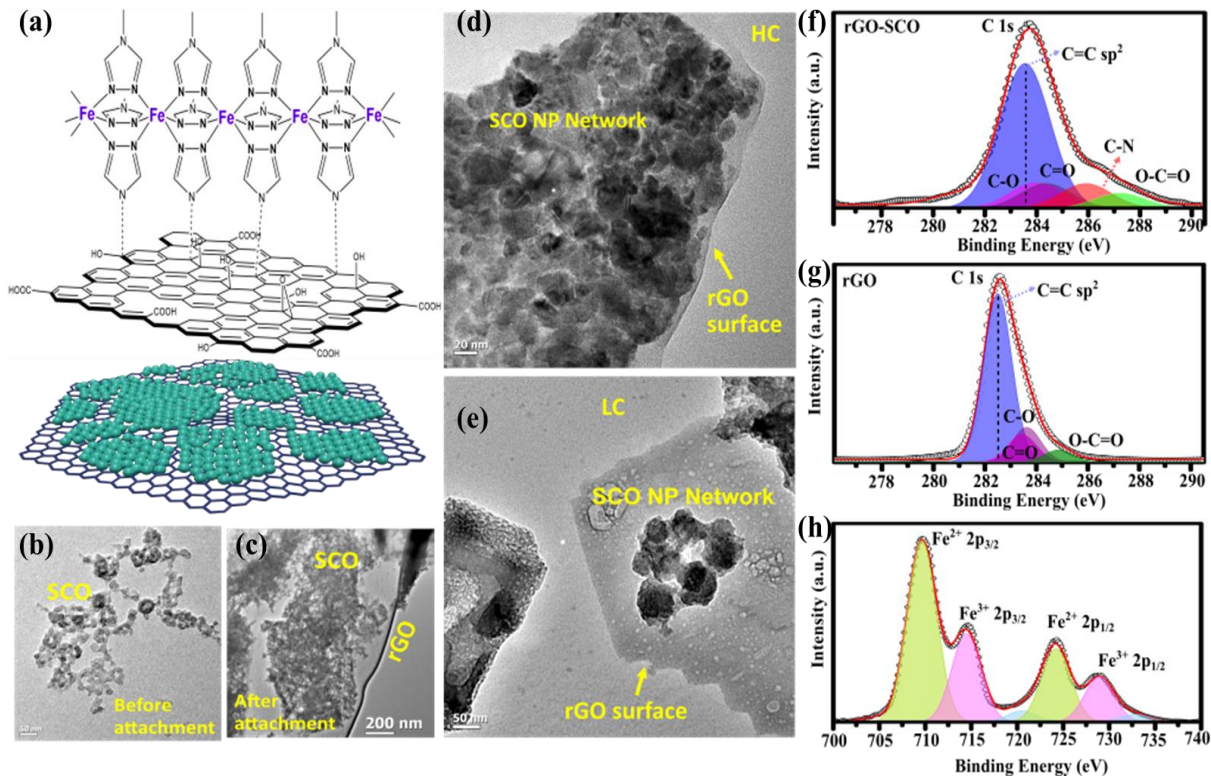

**Figure 1.** (a) Schematic representation of the SCO-rGO hybrid heterostructure. High resolution TEM images of (b) SCO nanoparticle, (c) SCO-nanoparticle attached to the rGO surface (large area), (d) high coverage (HC) sample and (e) low coverage (LC) sample. (f), (g) High resolution C 1s XPS spectra of SCO-rGO composite and pure rGO respectively. Deconvolution shows different bonding states like C-N bonding (red) appears in the hybrid. Shift in the binding-energy with broadened FWHM due to charge transfer is observed in SCO-rGO heterostructure. (h) XPS spectra for the oxidation states of Fe in Fe-containing Triazole complex.

**Magnetic measurements:**

**Magnetic field induced excited spin state transition in the hybrid nanostructure**

To understand the role of surface on the spin-state transition of the two separately prepared 2D SCO networks over the rGO surface, we have measured the dc mass susceptibility of pure Fe-Trz SCO and composite samples in a SQUID magnetometer (Quantum Design MPMS XL7) over the temperature range 240-400K in the heating-cooling mode at a rate 3 K/min under different applied magnetic fields (0.2 T, 1 T, 3 T, and 5 T). **Fig. 2a, c** compares the χT vs. T plots for HC and LC. At the lowest applied field of 2000 Oe (Fig. 2a-b), a gradual spin transition occurs for HC with a narrow hysteresis loop, representing weak cooperativity among the spin-active Fe(II)-centers in the 2D assembly of nanoparticles. Surprisingly, as the magnitude of the field is increased (i.e., from 1T to 5T), the transition occurs more sharply, along with a notable shift in the spin transition temperature ($T_{1/2}$), leading to broader hysteresis loops. This effect is more prominent in case of LC rather than HC. For LC, the sharpness (*aka* abruptness) of the transition and the hysteresis loop width are more tunable than HC. From the derivative plot of χT vs. T (**Fig. 2b, d**), we estimate precisely the $T_{1/2}^{up}$ and $T_{1/2}^{down}$, representing the characteristics spin transition temperatures during LS to HS and HS to LS transition, respectively. Moreover, the mean transition temperature $<T_{1/2}> = (T_{1/2}^{up} + T_{1/2}^{down})/2$ shifts towards room temperature in the SCO-rGO composite phases (**Fig. 2e**). While pure [Fe(HTrz)$_2$(Trz)](BF$_4$) phase has $<T_{1/2}>$ =362K; it reduces to 337K and 323K for LC and HC, respectively. For estimation, the difference between spin transition temperatures $T_{1/2}^{up}$ and $T_{1/2}^{down}$, ΔT(K), is plotted as a function of B (**Fig. 2f**).

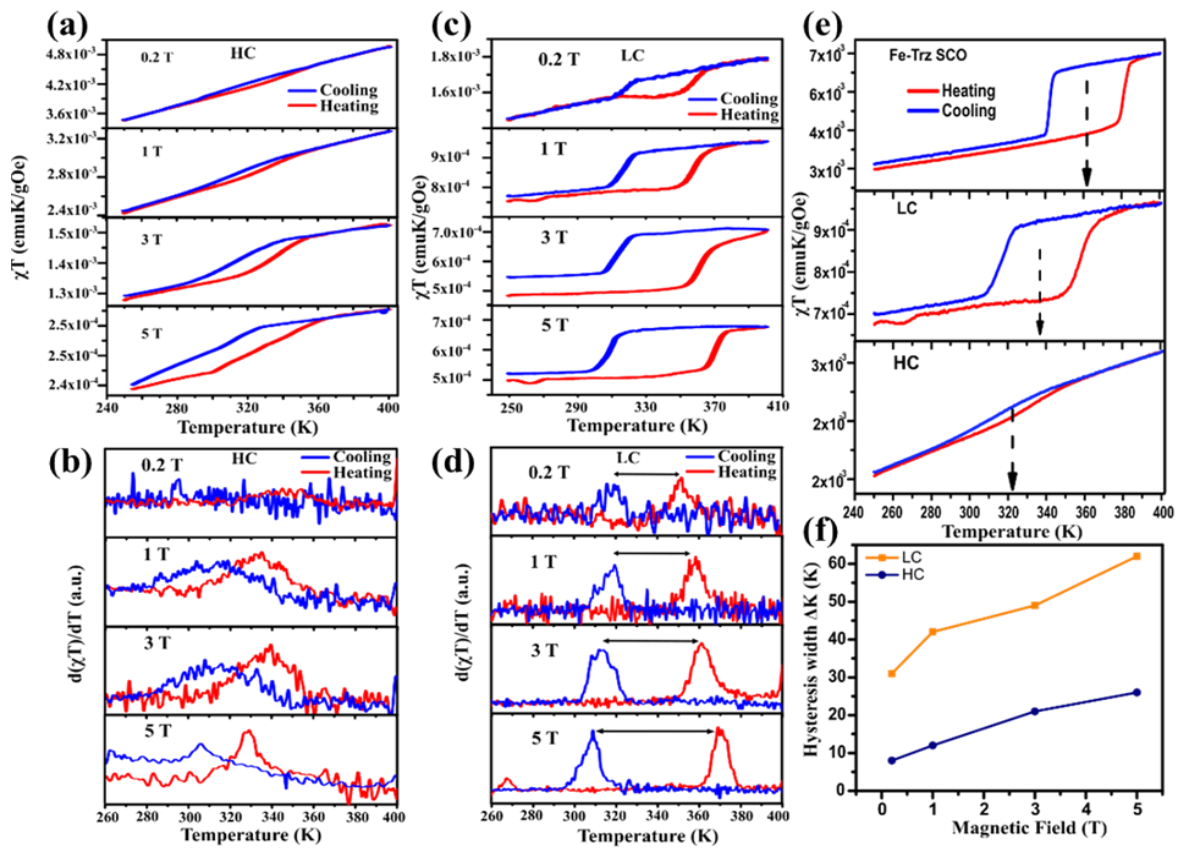

**Figure 2.** (a),(c) χT vs. T plots for two different composites with varying concentration of SCO shows field induced cooperative effect. (b), (d) Corresponding derivative of χT plot shows the shift in spin transition temperature with enhanced interaction. (e) The variation of middle of hysteresis, for pure SCO NPs, and LC SCO-rGO and HC SCO-rGO composites is represented. (f) The variation of hysteresis width with magnetic field for different concentration SCO-rGO composites

The maximum hysteresis width (ΔT) is ~26K and ~62K for HC and LC, respectively, whereas for pure Fe-Triazole nanostructures, it is ~39K.

In the earlier studies on pristine SCO compounds, it was observed that transition temperature could only be shifted a little based on different parameters like nanoparticle shape/size, heating-cooling rate, ligand geometry and chemical variation, synthetic conditions, etc.;[29,30] albeit, remains invariant under the magnetic field. Here, firstly, the shift of the transition temperature towards lower temperature in the hybrid heterostructure essentially attributed to the magnetic-field induced thermodynamic stabilization of the Fe(II) HS state in the SCO nanostructures arises by reducing the zero-point energy difference ($\Delta E^0_{HL}$, and its inhomogeneous distribution) towards negative values without sacrificing the cooperativity due to growth of SCO nanostructures on 2D rGO surface. Secondly, the enhanced sharpness of the hysteresis loops with the magnetic field indicates a field-induced spin-state transition. Finally, the broadening, as well as the sharpness of the hysteresis in the case of LC, further points towards the magnetic-field driven and the interface-induced enhanced cooperativity among the spin-active Fe(II)-centers in the 2D assembly of nanoparticles of the SCO-rGO composites. Our calculations, discussed later, further indicate the strengthening of the intra-chain Fe-Fe antiferromagnetic (AFM) interaction in SCO-rGO composites due to creation of the additional superexchange pathways through rGO.

Although the cooperativity leading to thermal hysteresis is believed to be predominantly elastic in nature, the magnetic exchange interaction between Fe centers in the chain adds onto this effect. On the heating path, one starts with the LS state of the Fe centers, and thus has no associated magnetic energy. On the other hand, in the cooling path, the Fe centers are in HS state and HS (S=2) -> LS (S=0) transition amounts to loss of magnetic exchange energy between Fe centers. This additional effect, which is present in cooling path, and not in heating path making $T_{1/2}^{down}$ less than $T_{1/2}^{up}$.

**Evolution of spontaneous magnetization in rGO/Fe-Trz nanocomposites:**

To understand the possible role of magnetic interactions and distinguish the role of SCO nanoparticles, we measure the zero-field-cooled (ZFC) – field-cooled (FC) susceptibility of both pure rGO and the SCO-rGO nanocomposite over the temperature range 2-400 K under 1T field at a heating-cooling rate of 3 K/min. For rGO (**Fig. 3a**), defect-induced paramagnetic states can be observed at low temperatures with no bifurcation between ZFC-FC modes, similar to the earlier reports in chemically synthesized rGO[35].

When the Fe-TRZ SCO network is attached to the rGO surface (in the case of LC), χ-T plots exhibit additional bifurcation in the ZFC-FC curves at a lower temperature, away from the thermal spin crossover regime (**Fig. 3b**), suggesting an exchange-induced effect in the hybrid. In HC (**Fig. 3c**), a broad bifurcation with susceptibility rising with temperature in the low temperature range is evident.

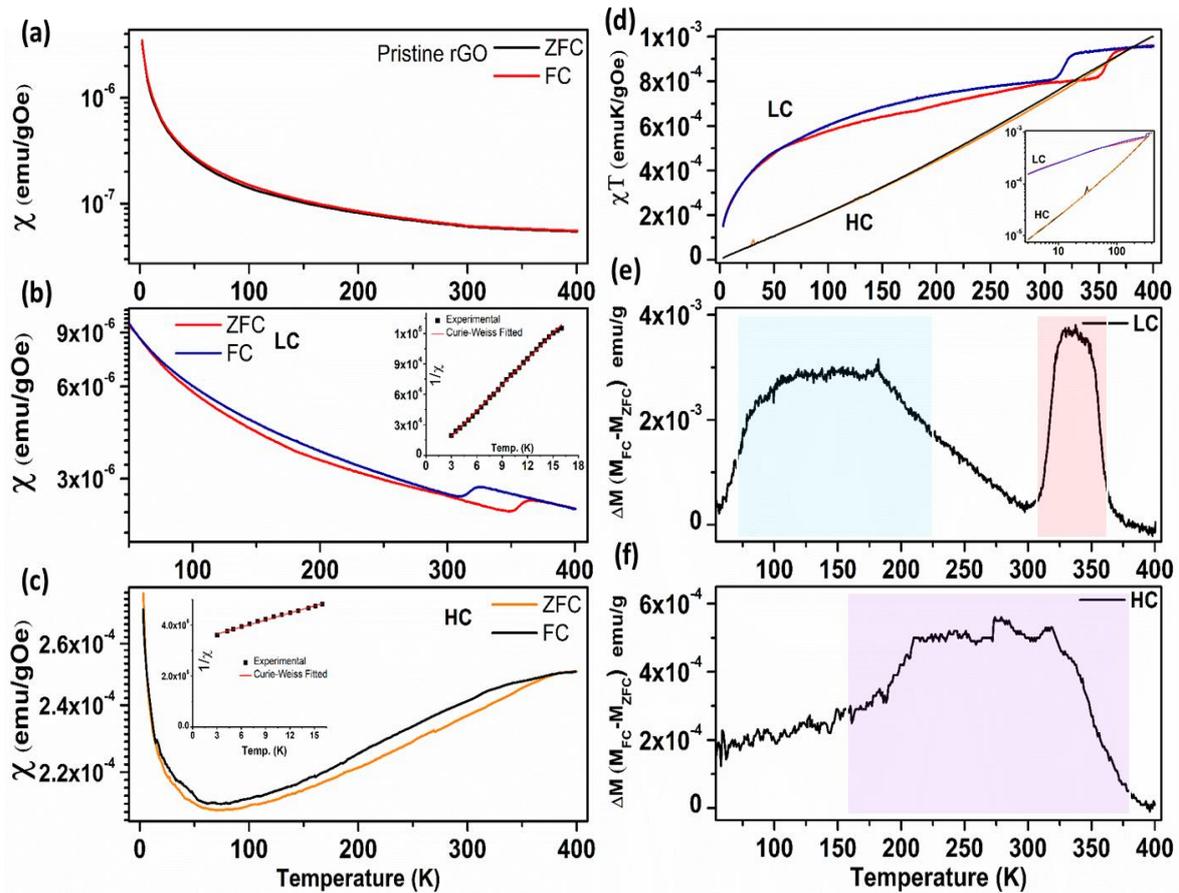

**Figure 3.** Mass susceptibility vs.Temperature plot for (a) rGO, (b) LC SCO-rGO composite, (c) HC SCO-rGO composite. Inset of (b) and (c) show the 1/ χ vs. T plot for respective hybrid system. (d) χT vs. T plot for the two composite systems. Inset shows the log-log plot for χT vs. T plot. (e) and (f) show the differential magnetization plot for LC and HC SCO-rGO composite respectively.

For further understanding of the magnetic interaction at low temperature, when $1/\chi$ vs. T is fitted with the modified Curie-Weiss law: $\chi = C/(T - \theta)$, a negative θ value of -36K is obtained for HC (**Inset Fig. 3c**) implying the presence of AFM coupling. While for LC, θ comes out as positive (~0.82 K), suggesting a weak ferromagnetic (FM) interaction. To compare the strength of the FM and AFM coupling for two different SCO coverages on rGO, we have plotted χT vs. T for LC and HC in **Fig. 3d**. For perfect antiferromagnets, when T→ 0 K, moment at 0 K should vanish. In this case, χT falls off more rapidly for HC than LC (inset of **Fig. 3d**) and reaches a magnitude of ~ $8.2\times10^{-6}$ at 2K, two orders of magnitude lower compared to LC (~$1.5\times10^{-4}$). It can be concluded that the AFM ordering in case of HC is much stronger than LC, due to thicker volume fraction of SCO nanoparticles network, further confirmed by DFT calculation, which is discussed later. In ΔM vs. T plot (**Fig. 3e, f**) for LC and HC, [ΔM = $M_{FC}$-$M_{ZFC}$], only a single broadened peak cantered around 264 K is observed for HC, whereas LC exhibits two distinct peaks at 136 K (broad) and 335 K (sharp). While the high-temperature peak (for LC) corresponds to the characteristic spin-crossover regime of Fe-Triazole, the additional low-temperature peak possibly corresponds to magnetic ordering in the complex as a result of interfacial hybridization.

**Appearance of bistable memory states in magnetization parameters:**

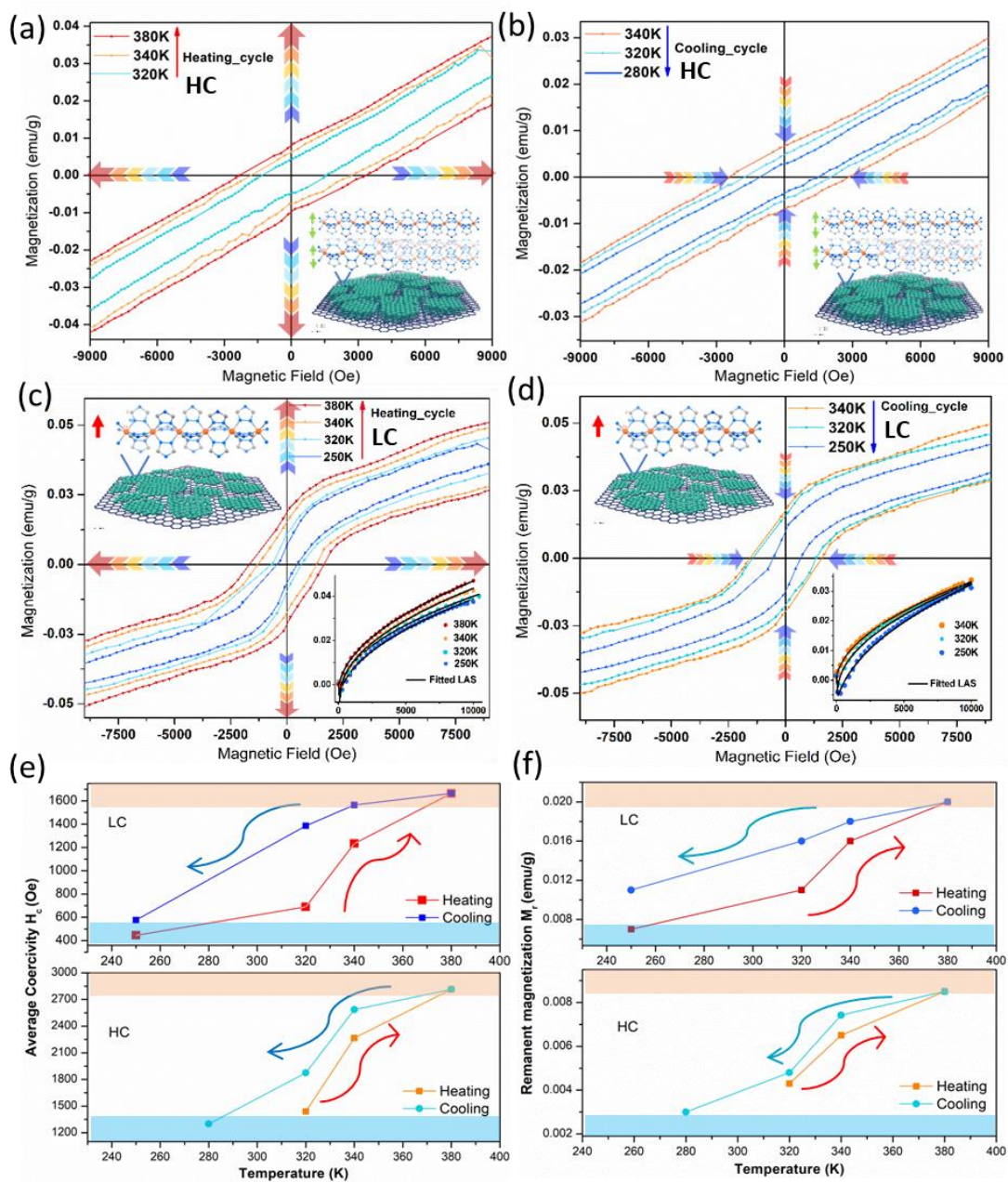

**Figure 4.** (a) and (b): Heating and cooling cycles hysteresis loops for sample HC. (c) and (d): Heating and cooling cycles hysteresis loops for LC. Insets: the fitting with law of approach to saturation magnetization. Also, the LC and HC are represented schematically in the corresponding insets. (e) Comparison of bistability in coercivity variation in the two hybrid (f) Comparison of remanent magnetization variation in two hybrid systems.

To trace the enhanced magnetic coupling among the Fe(II)-SCO centers during LS/HS transition, we have thoroughly collected the MH loop of Fe-Trz SCO/rGO hybrid at different temperatures in both heating and cooling cycles. In both cases, truncated hysteresis loops without complete saturation are noticed, similar to the observations in granular magnetic systems.[36][37] For HC, during heating mode at temperatures 320, 340, and 380 K, the hysteresis loop is widened, and the area under the curve increases compared to its previous temperature state (**Fig.4a**). This is unusual for magnetic systems, as with the rise in the thermal energy, anisotropy decreases in a much faster rate than magnetization[38][39]. Quantitatively, the anomalous giant increment in coercivity and remnant magnetization with temperature are 95% (1439 Oe at 320K → 2815 Oe at 380K) and 97%, respectively. This change is quite astonishing compared to pristine Fe-Trz SCO, where LS to HS transition involves diamagnetic to paramagnetic phase change with zero coercivity[40]. Contrastingly, in SCO-2D hybrid, a giant coercive field (~2800 Oe) and high exchange bias predict the presence of both ferromagnetic and anti-ferromagnetic coupling in the hybrid. During cooling mode hysteresis from 340K to 280K, the loop quenched to its initial state as the HS state depopulated to the LS state (**Fig. 4b**).

For LC, starting from 250K, the MH loop shows a considerable coercivity (445 Oe) and remnant magnetization (0.007 emu/g). During the heating cycle from 320 to 380K, the hysteresis loop is also widened here with a better tendency to saturate (**Fig. 4c**). At 380K, a maximum coercivity of 1664 Oe has reached with the highest remnant magnetization (0.02 emu/g). Hence, almost a 274% increment in the coercive field and a 185% increment in remnant magnetization have been observed in LC. During cooling from 340 to 250K, it is noticed that the decreasing trend in coercivity and remnant magnetization reversed as the high spin states start to depopulate (**Fig. 4d**). In the inset of Fig. 4c,d the schematic diagrams show a thin layer of SCO over rGO. To understand the dependency of magnetization with the applied field, we fit the initial growth curve with the Law of approach to saturation magnetization (LAS) which is also followed for typical magnetic systems[41]

$$M = M_s \left\{1 - \frac{a}{H} - \frac{b}{H^2}\right\} \chi H \quad \ldots\ldots\ldots\ldots\ldots\ldots\ldots\ldots\ldots\ldots\ldots\ldots \quad (1)$$

where $M_s$ is the saturation magnetization and $a$, $b$ are different constants. The $\chi H$ term refers to paramagnetism, and in some cases, it is found to be proportional to $H^{1/2}$ (i.e., our case)[42]. This term is added since the magnetization does not saturate completely due to presence of itinerant antiferromagnetic spins and consistent fitting is observed.

To arrest the newly evolved memory in magnetization parameters (due to bistability) in the spin states we plot coercivity and remnant magnetizations as a function of thermal cycle (**Fig. 4e,f**). The cyclic curves create a distinct hysteresis loop. The area under the loop in LC is larger than HC, suggesting stronger cooperativity among the Fe-centers. To check whether the enhanced ordering in the nanocomposite is due to defect-induced magnetism of rGO or not, we also compare the hysteresis of pristine rGO and SCO-rGO hybrid in the low-temperature range 2-50 K (**Supplementary information**

**Fig. S5**). For pure rGO, an explicit diamagnetic nature was observed due to the absence of any order, while for the hybrid, the MH curves follow the Brillouin type dependence,[43] discarding the role of any defect-induced magnetism due to rGO. The magnitude of the moment also increases largely compared to pure rGO, which is due to the addition of Fe(II) SCO centers in the 2D matrix. Details are given in the supporting information.

**Molecular field approach for evaluating coupling constants:**

In the SCO system, the nearest spins interact with Ising type spin Hamiltonian $H = -J_{ex,ij} S_i . S_j$. We have used molecular field approach to express the total susceptibility ($\chi_M$ as sum of three field components, two from the Fe HS/LS sites ($H_{Fe}$) and one from rGO interface ($H_G$). Field components ($H_{ij}$) are function of field coefficients ($n_{ij}$) and sublattice magnetization ($M_{Fe}$). While the field coefficients are governed by exchange coefficients $J_{ex,ij}$, the sublattice magnetizations have temperature dependence of Curie-Weiss type. After evaluating the mutual relationships among these parameters (see **Supplementary Information for details**), ultimately exchange coupling constant $J$ can be related to transition temperature $T_c$ by the following relationship,

$$T_c = \frac{\left(J_{ex,Fe_{HS}G}\right)\left[Z_{Fe_{HS}G} Z_{GFe_{HS}} S_{Fe_{HS}}\left(S_{Fe_{HS}}+1\right).S_G(S_G+1)\right]^{1/2}}{3k_B} \quad \ldots\ldots\ldots\ldots\ldots\ldots(2)$$

Here $S_i$ are the spin quantum number ($S_{Fe_{HS}} = 2, S_G = 1/2$), $Z_{ij}$ is the number of nearest neighbours which is 2 in Fe-Trz SCO system. For HC, the transition temperature obtained during Curie-Weiss fitting has been used in this equation that gives a negative exchange constant: $\frac{J_{HC,Fe_{HS}G}}{k_B} = -25.4$. While, in the case of LC, this gives a weak positive exchange constant: $\frac{J_{LC,Fe_{HS}G}}{k_B} = 0.58$. Hence, the negative coupling constant of HC signifies antiferromagnetic interaction (anti-cooperative), while positive exchange coupling constant for LC represents weakly ferromagnetic type (cooperative) interaction. Their relative magnitudes indicate the strength of interaction among the hybrid system.

**Transport properties**

**A. Sensing spin state dependency in conductance for the hybrid and pristine phases:**

To investigate the modulation in the electronic structure of rGO by surfaced SCO nanostructures, we perform temperature-dependent DC transport measurements of pure SCO and SCO-rGO hybrid heterostructures by fabricating two terminal devices on 300 nm Si/Si$^{++}$ substrate. As expected, pure Fe-Trz nanoparticles exhibit extremely low conductance (~10$^{-12}$ S at an applied bias of ~40V), i.e., in the insulating phase; however, the thermal spin crossover region is detected. While during the heating cycle (red curve), LS to HS transition results in an increment in conductance with a broad transition (**Fig. 5a**), cooling cycles (blue curve) display a drop in the conductance during the HS to LS jump. From a device perspective, pure SCO nanoparticle is unsuitable due to their extremely low conductance. Measuring transport under high bias (100V) leads to sample degradation[44][14]. The typical insulating nature of the pure SCO can be broken by embedding them on highly conducting rGO. Hybrid heterostructures have higher conductivity under low bias voltage with sample integrity [24].

To achieve better conductivity, we investigate the conductance behavior of the SCO-rGO composites under a thermal cycle. As expected, the conductance level of the SCO-rGO hybrid increases by seven orders of magnitude due to the presence of conducting rGO matrix. **Fig 5b** shows the 1$^{st}$ cycle of LC, where during the heating mode, the LS to HS state transition occurs sharply at around $T_{1/2}^{up}$ = 367 K, which gives a 136% drop in the conductance. During cooling mode, the reverse transition (HS to LS) occurs at around $T_{1/2}^{down}$ = 341 K with a 130% rise in conductance. Electrical hysteresis associated with two sharp changes in conductance can be related to the ON/Off state with a high memory channel—the hysteresis loop width, ΔK= 26 K for LC. For HC (**Fig. 5c**), the LS to HS transition occurs at $T_{1/2}^{up}$ = 393 K, with a 112% fall in conductance. During cooling mode, the reverse transition occurs at $T_{1/2}^{down}$ = 375 K, with a 110% rise in conductance and ΔK of 18 K. Despite hybridization to alien 2D rGO template, each sample reproduces large thermal hysteresis, which implies preservation of cooperativity even in the composite phase.

The hysteresis width (ΔK) of LC is almost 60% higher than HC, unlike the pristine SCO nanoparticles, where the hysteresis width decreases with the decrease in the volume of SCO [45–48]. The surface-to-volume ratio increases significantly for LC, while for HC, the out-of-plane 3D growth possibly masks the interfacial interaction [49]. The wider hysteresis loop and the shift in transition temperature in LC may be ascribed to the charge transfer-induced interfacial cooperativity enhancement and a variation in the $\Delta E^0_{HL}$, and its inhomogeneous distribution. Constricting the assembly of nanoparticles to the surface of rGO, the electrical pathway is preferentially aligned along a particular plane. Along this direction, the relative change in metal-ligand bond length would be the largest during LS/HS transition [50][51]. Additionally, to get a better understanding of the two terminal device architecture, a scheme is shown in **Fig. 5d**.

**B. Modulation in energy barrier during spin transition states mapped in transport study:**

In order to govern the spin state dependency on electrical conductivity, we consider a thermally activated process in the heterostructure rather than tunnelling as proposed by Constantin et.al. [52]. While it is difficult to pinpoint the exact hopping mechanism in the case of SCO-rGO hybrid due to the limited temperature range and relatively small conductance change, to get a qualitative idea about the variation in thermal activation energy during LS/HS transition, we fit the temperature-dependent conductance data using Arrhenius equation $\sigma_{dc}(T) = \sigma_0 exp(-E_a/k_B T)$, where σ$_0$ is the pre-exponential factor, $k_B$ is Boltzmann's constant, and E$_a$ is the activation energy. From **Fig. 5e** and **5f**, it is revealed that there is a significant change in the activation energy once the spin state varies from the LS to HS and vice versa. In LC, during the 1st cycle, $\langle E^a_{LS}\rangle=0.212$ eV rises to 0.714 eV $\langle E^a_{HS}\rangle$ for the HS state. While for the HC sample (1st cycle), $\langle E^a_{LS}\rangle=0.076$ eV, upon transiting to HS state, average activation energy rises to $\langle E^a_{HS}\rangle=0.239$ eV.

Considering transport by hopping mechanism the change in dc conductance can be related to modulation in hopping distance with a hopping frequency corresponding to the relevant phonon frequency. In the HS state, the molecular structure of SCO relaxes such as metal-ligand bond length increases. This leads to a shift in vibrational density of states to lower frequencies[53]. Hence, the phonon contribution is more relevant in the LS state rather than HS state, leading to higher conductance in the LS state. If we consider phonon frequency to couple the dc conductivity with the hopping frequency, Einstein's diffusion relation can be written following[54]:

$$\sigma_{dc} = \frac{n(el)^2}{6k_BT}\nu_P = \frac{n(el)^2}{6k_BT}\nu_{0P}\exp(-E_a/k_BT) \quad \text{...........................(3)}$$

where $n$ is the carried density, e is the charge, $l$ is the hopping distance, $\nu_P$ is hopping frequency, $\nu_P$ is the phonon frequency and $E_a$ is the activation energy[54]. Comparing this to the Arrhenius equation, we have found that the pre-exponential factor is basically consisted of hopping distance and hopping frequency. Thus, the observed conductance change is primarily related to the competition between hopping distance and hopping frequency in the LS and HS states. In the HS state, the hopping distance is more considerable due to higher activation energy and in the LS state, hopping frequency is higher, in line with the previous reports[44].

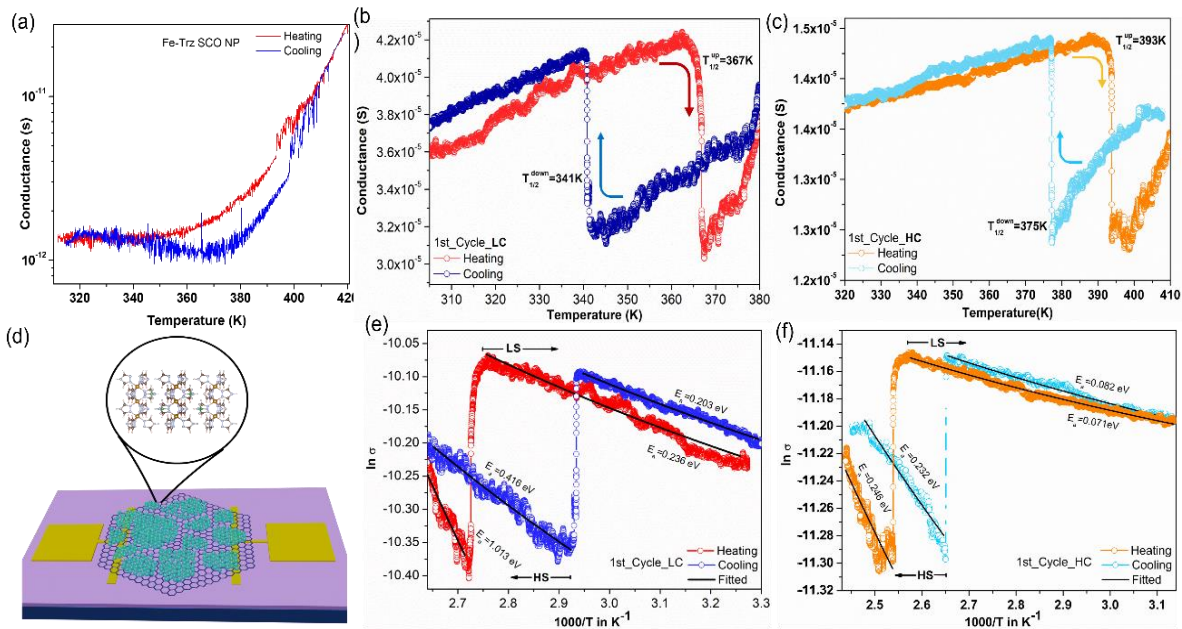

**Figure 5.** Thermal cycling in conductance measurement, (a) for pure SCO NPs, (b) 1st cycle of LC, (c) 1st cycle of HC, (d) Schematic of device architecture, (e) Fitting of Arrhenius equation in the 1st cycle of LC, and (f) in the 1st cycles for HC.

**DFT calculation of the SCO-rGO hybrid nanostructure:**

**A. Stability and Binding Energy calculations**:

To check the stability of the SCO-rGO structure, we first consider the binding of SCO polymeric chain with rGO as well as with graphene. $E_{binding}$ with rGO as substrate (-1.15 eV) turns out to be much lower compared to that with graphene (-0.25 eV). This indicates that embedding of SCO complex on rGO compared to graphene substrate is more stable. Further, there are two free ends in Fe-triazole chain available to attach on rGO surface, H atom of H-trz ligand, and N atom of trz$^-$ ligand, leading to two possible configurations of the model system, configuration-1 and configuration- 2, respectively (cf **Fig. 6a, b**.). $E_{binding}$ of configuration-1 is found to be 18 meV lower than that of configuration-2. We further note that the distance between the free end of the chain and rGO increases in configuration-2 during structural relaxation resulting in the rGO surface curving towards the chain compared to configuration-1 which is evident from **Fig. 6a, b**. Hence, the interaction between SCO NPs and rGO surface is expected to be more in configuration-1 than configuration-2. In the following, we restrict our discussion to configuration-1 only.

**B. Choice of U and Optimized Structure**

A crucial parameter in DFT calculation is the choice of Hubbard U, which is instrumental in faithful description of the SCO behavior. The choice of U in the present calculations is guided by the thermal transition temperature $T_{1/2}$ of SCO compound [Fe(Htrz)$_2$(trz)](BF$_4$). It is possible to have an experimental estimate of electronic energy difference between HS and LS states of the SCO compound $\Delta E_{HL}$ (= $E_{HS}$ - $E_{LS}$) using following relation $\Delta E_{HL} = \Delta S_{HL} T_{1/2}$ where $\Delta S_{HL}$ is the entropy difference and found to vary within small range (~ 5 cm$^{-1}$K$^{-1}$)[16]. SCO compound favors the LS state at low temperature ($\Delta E_{HL}$= +ve) and as temperature increases, it favors the HS state ($\Delta E_{HL}$= -ve), leading to entropy-driven thermal spin transition. We have computed total energy E for both HS and LS structures taking series of U values ranging from 0 to 5 eV and computed $\Delta E_{HL}$ as a function of U as plotted in **Fig. 6c**. $\Delta E_{HL}$ changes sign around U = 2.5 eV, and hence the correct ground state, i.e., HS state is captured. Experimental estimate of $\Delta E_{HL}$ (= -0.22 eV) is found in good agreement for U = 2.6 eV, which is used for subsequent calculation.

**Fig. 6d** shows the detailed optimized structure of the model system, as calculated within GGA+U formulation with choice of U = 2.6 eV, as discussed above. Carbon atoms, C37 and C38 of rGO are attached to -OH groups O1-H17 and O2-H18 of SCO molecules, respectively. During the structural relaxation, they move slightly upwards. Hence the distance between the free end of Fe-triazole chain and rGO, i.e., N15-H18 and H16-O1, decreases and becomes 1.82Å and 1.70Å, respectively. This gives rise to two potential charge transfer paths between SCO and rGO at the interface.

## C. Charge transfer analysis

**1. Charge density difference plot:** To investigate the charge transfer between SCO NPs and rGO, we have computed the charge density difference ($\Delta\rho$). $\Delta\rho$ can be defined as: $\Delta\rho = \rho_{model} - \rho_{chain} - \rho_{rGO}$, where $\rho_{model}$ denotes the charge density of the model system containing Fe-triazole chain embedded on rGO. $\rho_{chain}$ and $\rho_{rGO}$ indicate the charge densities of Fe-triazole chain and rGO, respectively. According to the formulation, +ve value of $\Delta\rho$ represents an accumulated charge, whereas -ve value denotes the depleted charge. **Fig 6e** shows the plot of $\Delta\rho$ isosurfaces for the model system. This clearly shows finite charge accumulation at Fe-triazole chain and rGO interfaces, one is between N15 and H18 atoms, and the other is between H16 and O1 atoms. Hence, charge transfer between SCO NPs and rGO can occur through their interfaces.

**2. Quantitative analysis of charge transfer: Bader charge calculation**

Next, to quantify charge transfer, we have performed bader charge difference of the model system on the individual atoms, Fe-triazole chain, and rGO separately. Bader charge difference can be obtained using the same formula as that of $\Delta\rho$. We found a finite bader charge on the interfacial atoms of the model system, and the sign of the bader charge matches that of $\Delta\rho$, supporting the charge transfer process. Bader charge on N15 and H18 atom are 0.025e and -0.069e, whereas H16 and O1 have charges -0.070e and -0.005e. Bader charge on carbon atoms C37 and C38 of rGO is found to be -0.021e and 0.041e, respectively. (**Table S2 in supporting information**).

## D. Spin-polarized density of states (DOS):

The computed spin-polarized DOS of the hybrid system with SCO molecules in HS geometry, is shown in **Fig. 6f**. Total DOS shows that there is an asymmetry in spin-up and spin-down electron density, indicating presence of net magnetic moment in the system. To investigate the source of this, we have further calculated the projection of the total DOS on the Fe 3d and N 2p orbitals associated with Fe-triazole chain and C 2p and O 2p orbitals of rGO. Since Fe(II) is in HS state (S=2), spin-up $t_{2g}$ and $e_g$ states are completely filled, whereas spin down $t_{2g}$ states are partially filled with one electron. Examination of the projected DOS, reveals the uncompensated magnetic moment is the system primarily arisies from Fe 3d orbitals, with induced magnetism at N 2p as well as C 2p atoms, due to finite hydrization between the two subsystems of SCO and rGO. The finite hybridization of SCO states with that of rGO has two important effects, a) it enhances the intra-chain magnetic exchange between Fe centres which is antiferromagnetic in nature, by opening up additional super-exchange pathways, and b) enables inter-chain interaction via RKKY type interaction between magnetic chains, which in principle can be also ferromagnetic, mediated by conduction electrons of rGO. **E. Magnetic exchange correlation**

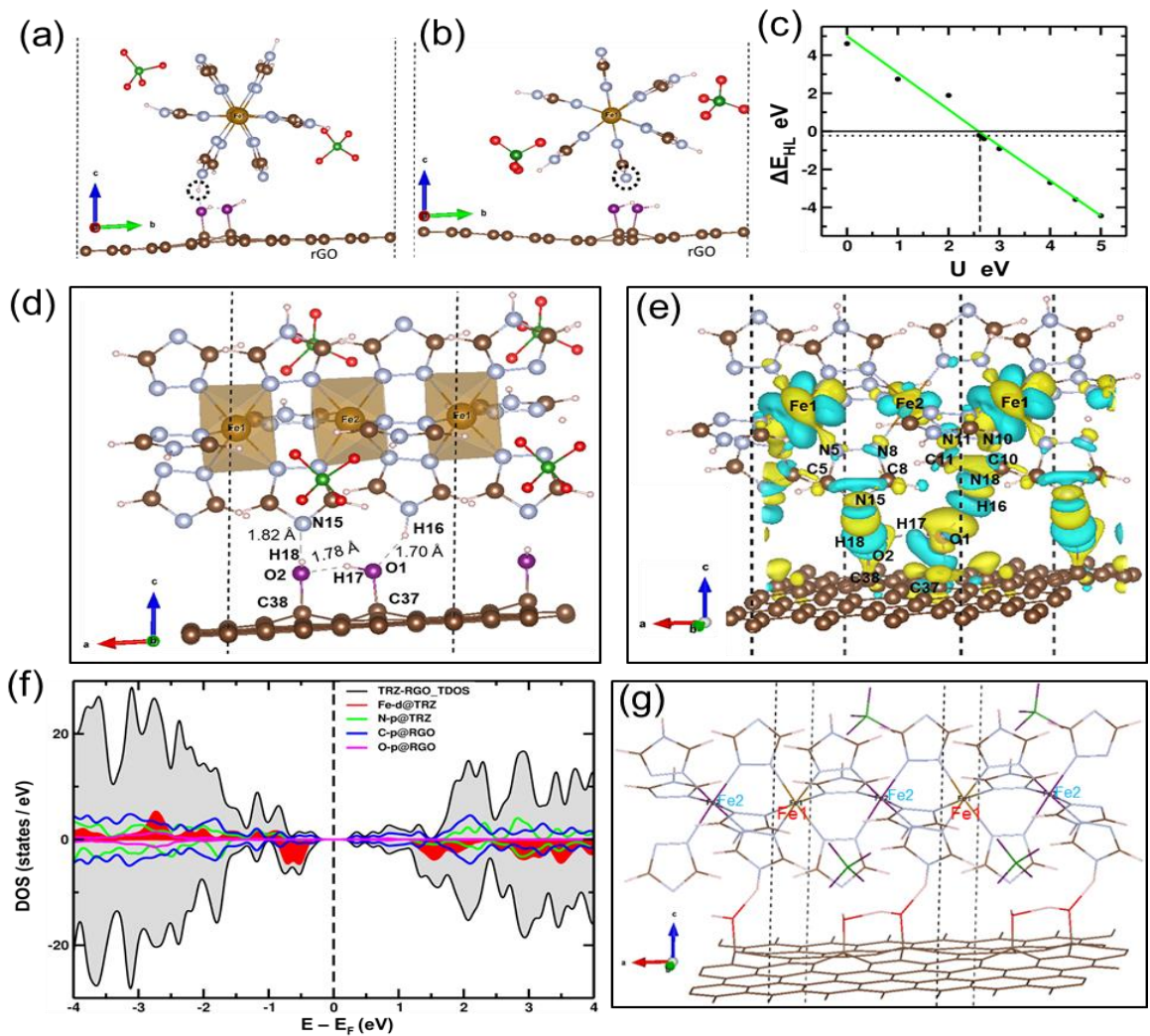

**Figure 6.** Model system containing single Fe-triazole chain embedded on rGO surface (a) configuration-1 and (b) configuration-2. Grey:N and brown:C  Pink: H, green: B and Red: F corresponding to counterion BF4-. (c) Calculated values of ΔEHL vs. U ε [0,5] with a linear fit. (d) Optimized structure of model system. Free ends of the Fe-triazole chain and -OH groups of RGO are marked and corresponding distances are mentioned. (e) Plot of the charge density difference Δρ. The isosurface value is set at 0.002 e-/Å3 with depleted and accumulated charges shown by cyan and yellow colored isosurfaces, respectively. (f) PDOS of Fe d, N p orbitals of Fe-triazole chain and C p and O p orbitals of rGO. Asymmetric total DOS is shown by black shaded area.  (g) Intrachain magnetic exchange interaction between Fe(II) sites of the model system.

There are two types of intra-chain interactions that can contribute to the spin transition of a SCO polymeric system. One is phononic elastic interaction and second is magnetic exchange interaction *J*. Second interaction is due to the nearest neighbour superexchange between the Fe(II) centers, which is typically of antiferromagnetic nature[16,55]. Now, to understand the influence of rGO on cooperativity among the Fe(II) centers affecting spin transition of SCO NPs, we compute intrachain superexchange coupling $J$ in presence and absence of rGO (**Fig. 6g**).

Considering a spin-Hamiltonian $H = JS_iS_{i+1}$ between nearest neighbors, Fe$^{2+}$ spins $S_i$, the ferromagnetic and antiferromagnetic energies for two Fe$^{2+}$ ions in the unit cell of the Fe-triazole network are given by $E_{FM} = 8J$ and $E_{AFM} = -8J$. $E_{FM} - E_{AFM}$ for model system is 43.2 meV while, for only Fe-triazole chain, $E_{FM} - E_{AFM}$ =36.8 meV . Thus a large increase in the intra-chain antiferromagnetic exchange is noticed in presence of rGO, supporting the expectation finite hybridization between SCO and rGO results in opening up of additional superexchange pathways. From this calculation, it is to be concluded that the role of rGO is to strengthen the superexchange interaction between Fe(II) centers that may lead to long-range magnetic ordering in the hybrid structure.

**Conclusion**

In summary, our results demonstrate that the fabrication of Fe-Trz-based SCO hybrids on 2D rGO not only retain the desirable properties of the two components, namely SCO and rGO, in presence of each other, but also get enhanced by helping each other, thereby overcoming the bottleneck for device application. The characterization of the designed device by spectroscopic tool as well as first-principle calculations, demonstrate the interfacial charge transfer between rGO and SCO. By tuning the coverage area of rGO by the SCO nanoparticle-networks, tunable magnetic property is achieved. While the low temperature behaviour of the susceptibility data for the high-coverage sample clearly indicates the dominance of anti-ferromagnetic coupling, for the low coverage case, a weak ferromagnetic nature is observed, suggesting an additional ferromagnetic interaction possibly induced by the formation of rGO/SCO interface. Appearance of temperature dependent hysteresis in the M-H curves with large coercive field around the spin crossover regime at higher temperature, further establishes the presence of both antiferromagnetic intra-chain interaction and ferromagnetic inter-chain interaction in the hybrid. The ab-initio calculations based on a simple model which is more applicable for the HC, verify the enhancement of the antiferromagnetic interaction between the Fe centres via the charge transfer with the conducting rGO substrate. The ferromagnetic interaction may arise due to the emergence of pinned magnetic moment in the conducting electron of rGO, providing memory hysteresis with large coercive field. We further fabricated micron scale device to measure the conductivity of the hybrid system which exhibits superior conductivity with bistability, showing promises towards memory application. Thus, in compared to pristine SCO NPs, the hybrid nanostructures show enhanced cooperativity, better conductance-bistability and a variation in the $\Delta E^0_{HL}$ and its inhomogeneous distribution that led to a shift in transition near room temperature for practical manifestation. The 2D rGO template gives a platform for stabilizing the enhanced coupling in spin centres of SCO network. Therefore, the SCO 2D-substrate interactions deliver the mechanical and conducting backbone for developing bistable NP-based hybrid devices.

**Acknowledgments:**

S.B. acknowledges Japan Society for the Promotion of Science for providing JSPS International postdoctoral fellowship. H.T. thanks JSPS for supporting via Grants-in-Aid for Scientific Research (KAKENHI). S.M. acknowledges the CSIR Fellowship program, Ministry of Science & Technology, Government of India, for providing a research fellowship (09/575(0118)/2017-EMR-I). A.N.P. acknowledges financial support from DST-Nano Mission (Grant No. DST/NM/TUE/QM-10/2019). We acknowledge the characterization facilities from the TRC project and the clean room fabrication facilities under the technical cell of S.N.B.N.C.B.S. We thank Prof. N. Sarkar, Department of Chemistry, IIT Kharagpur, for providing the DLS measurement facility. We also thank the Department of Chemistry, IIT Kharagpur for the FTIR facility. We are thankful to the SEM facility, Department of Chemical Engineering, IIT Kharagpur, for the EDX analyses. C. D. acknowledges IIT Kharagpur for the research fellowship. P. C. gratefully acknowledges the financial support from SERB (Grant no. ECR/2018/000923) and IIT Kharagpur (Grant no. IIT/SRIC/CY/ENE/2018-19/194). T.S.D acknowledges J. C. Bose National Fellowship (grant no. JCB/2020/000004) for funding.

# Supporting Information

# Establishing Magnetic Coupling in Spin-crossover-2D Hybrid Nanostructures via Interfacial Charge-transfer Interaction

Shatabda Bhattacharya[1, 3], Shubhadip Moulick[1#], Chinmoy Das[2#], Shiladitya Karmakar[1#], Hirokazu Tada[3], Tanusri Saha-Dasgupta[1], Pradip Chakraborty[2]*, Atindra Nath Pal[1]*

[1]*Department of Condensed Matter Physics and Material Sciences, S. N. Bose National Centre for Basic Sciences, Salt Lake City, Kolkata - 700 106, India*

[2]*Department of Chemistry, Indian Institute of Technology Kharagpur, Kharagpur-721302, India*

[3]*Department of Materials Engineering Science, Graduate School of Engineering Science, Osaka University, Toyonaka, 560-8531, Japan*

*Corresponding Author: atin@bose.res.in; pradipc@chem.iitkgp.ac.in

(# contributed equally to this work)

**Materials**

Precursor materials, $Fe(BF_4)_2 \cdot 6H_2O$ (Iron(II) tetrafluoroborate hexahydrate) and 1,2,4-triazole (Htrz), ascorbic acid, Dioctyl sulfosuccinate sodium salt as surfactant (NaAOT); behenic acid as co-surfactant and n-octane were purchased from Sigma Aldrich and used for the synthesis without any further purification. Solvents used in the synthesis were of laboratory reagent grade.

**Color change due to thermal spin state transition**

The left picture of Fig. S1 shows the purple-colored (i.e., in the low-spin state) ethanolic nanoparticle suspension at room temperature. Upon heating to 385 K, the nanoparticle suspension becomes transparent and colorless (shown in the right picture of Fig. S1), indicating the switching of the population to the high-spin state. Upon cooling again to 340K, the purple color reappears due to the thermal population returning to the low-spin state, indicating the reversible switching between the two electronic spin-states as a function of temperature (i.e., the thermal spin transition phenomenon).

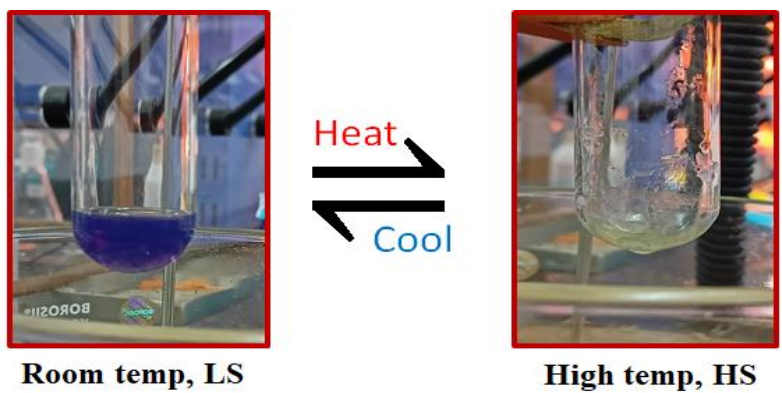

**Fig. S1.** Reversible color change due to heating/cooling shows thermal spin state transition

**Particle size determination through DLS analysis**

The size distribution (shown in Fig. S2) of the as-synthesized nanoparticles was measured using the Malvern Nano ZS instrument consisting of 4mW He-Ne laser ($\lambda$ = 632.8 nm). The detector angle was set to 173°.

As shown in Fig. 2, the size distribution of the nanoparticles is in the range of 30-40 nm. After the Gaussian fit of the size distribution (using software OriginPro 9), the average size of the nanoparticles is 34.87 nm

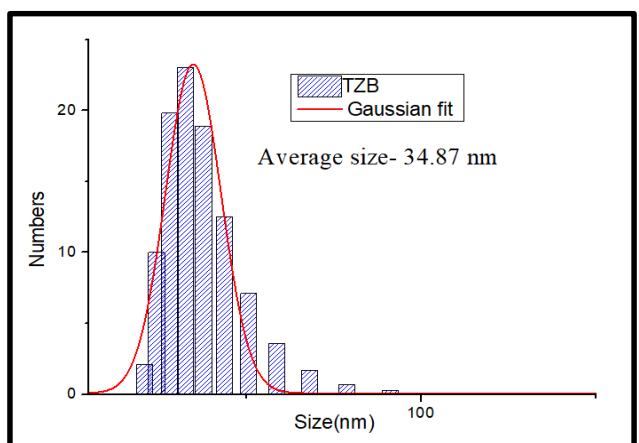

**Fig S2.** DLS plot of pure [Fe(Htrz)$_2$(trz)]BF$_4$

**FTIR analysis of pristine [Fe(Htrz)$_2$(trz)]BF$_4$**

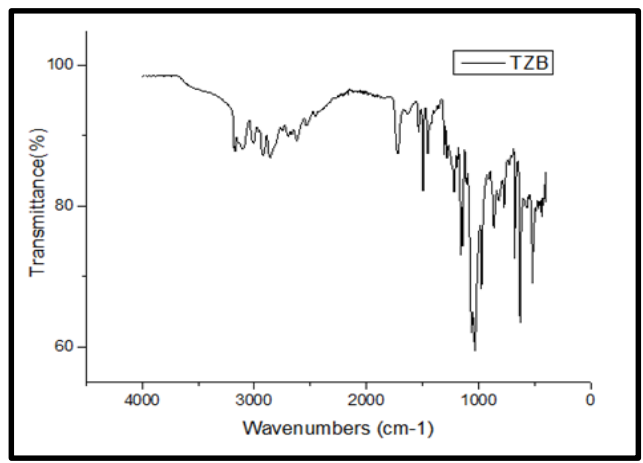

**Fig S3.** FTIR spectrum of the desired [Fe(Htrz)$_2$(trz)]BF$_4$

**Table S1.** Energy-dispersive X-ray analysis of the effective composition of C, N, Fe and F (in wt. %) at three different randomly chosen sites

| Compound | | C (%) | N(%) | Fe (%) | F (%) | Formula |
|---|---|---|---|---|---|---|
| Site 1 | Calc | 27.30 | 31.84 | 14.10 | 19.19 | [Fe(Htrz)$_2$(trz)]BF$_4$ $C_9H_{19}FeN_9BF_4$ Mol. Wt.: 395.957 |
| | Exp | 26.80 | 29.49 | 9.33 | 20.39 | |
| Site 2 | Calc | 27.30 | 31.84 | 14.10 | 19.19 | |
| | Exp | 24.59 | 28.10 | 12.09 | 23.00 | |
| Site-3 | Cal | 27.30 | 31.84 | 14.10 | 19.19 | |
| | Exp | 26.49 | 30.82 | 10.90 | 21.04 | |

**Energy dispersive X-ray (EDX) spectrum**

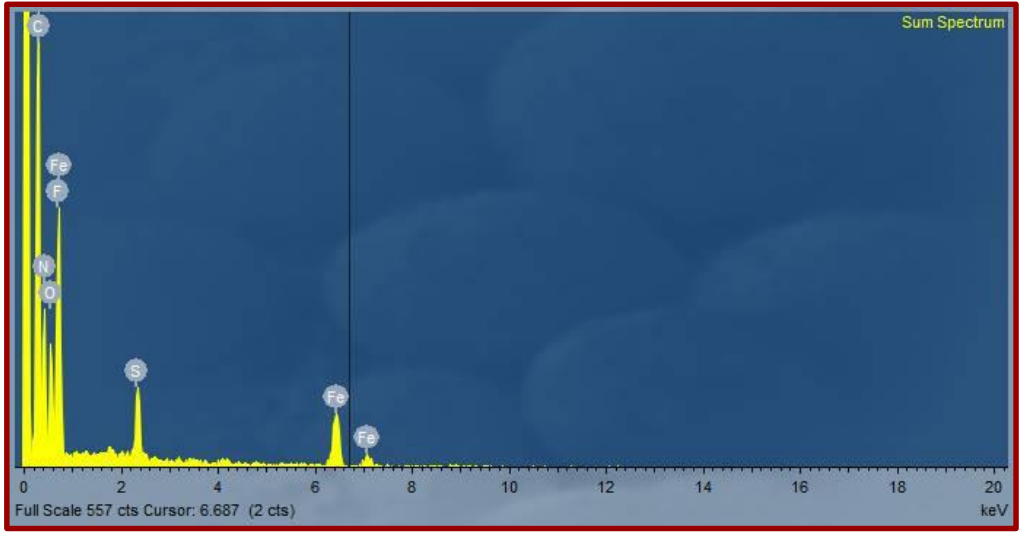

**Fig. S4.** EDX spectrum of the [Fe(Htrz)$_2$(trz)]BF$_4$ nanoparticles, where major constituent atoms (*i.e.* C, N, Fe and F) are shown. S and O contribution as shown in the spectrum are coming from the residual surfactant present in the washed and aged nanoparticles. H and B (not shown) are below the threshold detection limit.

**Comparison of MH loops at low temperature for pure rGO and rGO/SCO hybrid**

In the lowest temperature 2 K, a weak paramagnetism is observed due to some defect states in chemically synthesized rGO, but from 10 K onwards, a clear diamagnetic nature started to reveal until 300 K, which is obvious due to absence of any ordering (**Fig S5a**). However, when SCO-2D assemblies are attached to the rGO surface, the magnetization nature changes significantly starting from low temperature 2 K. For SCO-rGO hybrid, in the low temperature 2-50 K, a non-saturating magnetization is revealed (**Fig. S5b**). The magnitude of the moment is also increased largely compared to the pure rGO, which is due to the addition of Fe(II) centers in the 2D matrix. We fitted the 2 K curve with Brillouin function as follows:

$$B_J(x) = ((2J+1)/2J) \coth((2J+1)/2J\, x) - 1/2J \coth(x/2J); x = (gJ\mu_B H)/(K_B T)$$

…………………
(1)

Antiferromagnetic coupled system shows Brillouin type dependence at very low temperatures. Since magnetization does not saturate at 2 K, there is a weak interaction among the Fe(II) LS spin centers at low temperature. Unless there is a spin transition to HS state, the interaction does not start. To get FM ordering in SCO, it is important to get a high number of interacting HS states. This is achieved in this case through substrate introduction (rGO) and via charge transfer at the interface.

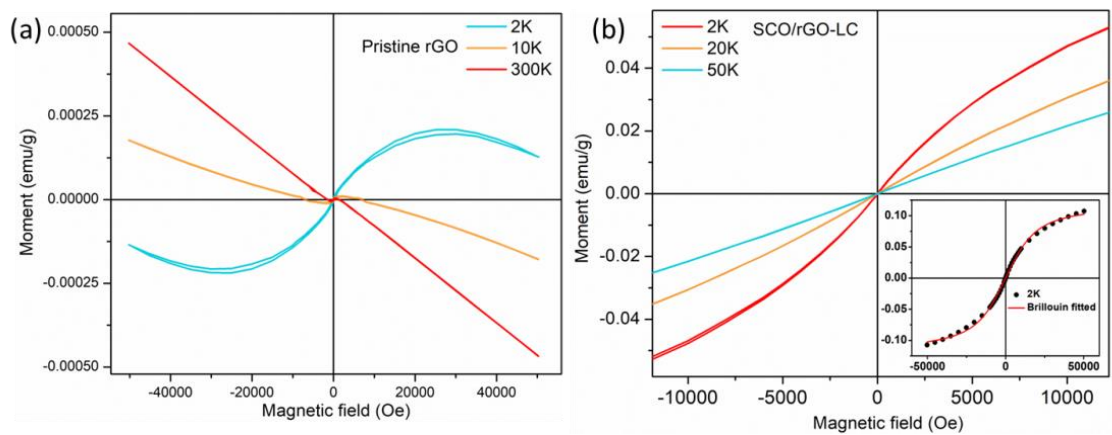

**Fig. S5.** Comparison of MH hysteresis loops at low temperature (a) pristine rGO, (b) SCO-2D composite

**Ab initio based density functional theory calculation**

**1. Simulation Model**

Ab-initio study on hybrid iron(II) based spin-crossover nanoparticle-reduced graphene oxide (SCO-rGO) composite is not computationally tractable because of its large size (~30-40 nm) and huge number of atoms. SCO NPs have been prepared from triazole based polymeric compound of the formula [Fe(Htrz)$_2$(trz)](BF$_4$) where Htrz: 1,2,4-triazole and trz$^-$ stands for deprotonated triazolato. Since the interaction between SCO NPs and rGO is mediated through their common interface, we build up a model system to simulate the interfacial effect between corresponding SCO compound and rGO.

**SCO compound**: The crystallographic data of Low Spin (LS) and High Spin (HS) structure of [Fe(Htrz)$_2$(trz)](BF$_4$) reported by Grosjean et.al.[2] are used as starting point to construct the model. The structure consist of Fe(II) centers with neighboring Fe(II)'s connected by a bridge made of two Htrz ligands and one trz$^-$ ligand, forming an one-dimensional chain-like network. This generates an octahedral environment of six nitrogen atoms surrounding each Fe(II) center. Each counterion BF$_4^-$ located in cavities formed by triazole ligands to maintain charge neutrality. The model aims to understand the influence of rGO on cooperativity among Fe(II) centers affecting spin transition of SCO NPs.

**The model:** In building up the model first we extract a single chain of Fe-triazole polymer having two units of [Fe(Htrz)$_2$(trz)](BF$_4$) inside unit cell from its reported crystal structure. Secondly, we construct a simulation cell using Latticematch method based on rotation matrix as implemented in LatticeMatch code[3] which ensures minimum lattice mismatch (~ 0.67%) between hexagonal graphene (s.g. P6$_3$/mmm) and orthrhombic [Fe(Htrz)$_2$(trz)](BF$_4$) (s.g. Pnma) unit cell. Inside the simulation cell, extracted [Fe(Htrz)$_2$(trz)](BF$_4$) chain is placed along *a* direction on top of suitably chosen supercell of graphene (*ab plane*). The in-plane lattice parameter of the simulation cell is as follows: a = 7.325 Å and b = 17.347 Å, which contains total 106 number of atoms out of which 58 is from the chain and 48 carbon atoms from graphene making our model computationally tractable for ab-initio study. Now in order to make rGO, two hydroxyl (-OH) groups are attached to the carbon atoms of graphene and adjusted right below the free ends of the chain to ensure greater interaction at the interface. Distance between free end atom of chain and rGO is set to ~ 2 Å.

## 2. Computational details

Ab initio calculation is carried out within the framework of density functional theory supplemented by hubbard U (DFT+U) approach employing projector augmented-wave pseudopotentials and Perdew-Burke-Ernzerhof generalized gradient approximation (PBE-GGA) [4] as exchange correlation functional as implemented in the plane-wave based Vienna ab initio simulation package (VASP) [5]. The cutoff energy of the plane-wave basis is chosen to be 500 eV, sufficient to achieve convergence in self-consistent field (SCF) calculations. To minimize the artificial interaction between model system and its images along the out-of-plane *c* direction within the periodic setup of the calculation, a vacuum space of ~ 16 Å is used. Two Fe-triazole chains are separated by 8 Å along *b* direction to avoid interchain interaction.

The relaxation of model structure is done with respect to internal atomic coordinates and the simulation cell volume using a convergence threshold of $10^{-5}$ eV for the total energy and 0.05 eV/Å for maximum force/atom employing 4x2x1 Monkhorst-Pack mesh[6]. More densed k mesh of size 8x4x1 is used for SCF calculation with increased energy convergence threshold of $10^{-8}$ eV. Converged SCF is used to compute density of states, charge density and bader charges on individual atoms of composite as well as component structures.

**Table S2:** Bader charge analysis of the individual atoms of the model system

| System: | Δρ in e |
|---|---|
| H18@rgo | -0.0697 |
| O2@rgo | -0.0277 |
| C38@rgo | 0.0412 |
| O1@rgo | -0.0047 |
| C37@rgo | -0.0204 |

**Molecular field approach for evaluating coupling constants:**

We have used molecular field approach to express $\chi_M$ as sum of three components. Molecular field components from Fe$_{HS}$, Fe$_{LS}$ and rGO moment can be expressed as:

$$H_{Fe_{HS}} = n_{Fe_{HS}G} M_G \quad , \quad H_{Fe_{LS}} = n_{Fe_{LS}G} M_G \quad , \text{ and } \quad H_G = n_{GFe_{HS}} M_{Fe_{HS}} + n_{GFe_{LS}} M_{Fe_{LS}}$$
………(2)

Here $H_{Fe_{HS}}, H_{Fe_{LS}}$ and $H_G$ are the molecular fields acting on Fe HS, LS and rGO sites and $n_{ij}$ are the field coefficients, and $M_{Fe_{HS}}$, $M_{Fe_{LS}}$ and $M_G$ are sublattice magnetization from different Fe and rGO sites respectively. Based on modified Curie-Weiss law, in paramagnetic region, these sublattice magnetizations have temperature dependence as follows:

$$M_{Fe_{HS}} = \frac{C_{Fe_{HS}}(H_0 + H_{Fe_{HS}})}{T} \, , \, M_{Fe_{LS}} = \frac{C_{Fe_{LS}}(H_0 + H_{Fe_{LS}})}{T} \text{ and } M_G = \frac{C_G(H_0 + H_G)}{T} \dots\dots\dots\dots\dots\dots$$
(3)

Here $H_0$ is the external magnetic field, $C_{Fe_{HS}}$, $C_{Fe_{LS}}$ and $C_G$ are the Curie constants of different sublattice components. Combining these sublattice magnetizations, $\chi_M$ can be defined as

$$\chi_M = \left(M_{Fe_{HS}} + M_{Fe_{LS}} + M_G\right)\Big/H_0 \quad \ldots\ldots(4)$$

The Curie's constant for different sites is related to spin magnetic moment as follows:

$$C_{Fe_{HS}} = \frac{N_A g^2 \mu_B^2 S_{Fe_{HS}}\left(S_{Fe_{HS}}+1\right)}{3k_B}, \quad C_{Fe_{LS}} = \frac{N_A g^2 \mu_B^2 S_{Fe_{LS}}\left(S_{Fe_{LS}}+1\right)}{3k_B}, \text{ and}$$

$$C_G = \frac{N_A g^2 \mu_B^2 S_G(S_G+1)}{3k_B}, \ldots\ldots(5)$$

The $n_{ij}$ (field coefficients) can be related to exchange coefficients $J_{ex,ij}$ through:

$$n_{Fe_{HS}G} = \frac{Z_{Fe_{HS}G}}{N_A g^2 \mu_B^2} J_{ex_{Fe_{HS}G}}, \quad n_{Fe_{LS}G} = \frac{Z_{Fe_{LS}G}}{N_A g^2 \mu_B^2} J_{ex_{Fe_{LS}G}}, \quad n_{GFe_{HS}} = \frac{Z_{GFe_{HS}}}{2xN_A g^2 \mu_B^2} J_{ex_{GFe_{HS}}} \text{ and}$$

$$n_{GFe_{LS}} = \frac{Z_{GFe_{LS}}}{2(1-x)N_A g^2 \mu_B^2} J_{ex_{GFe_{LS}}} \quad \ldots\ldots(6)$$

Here $Z_{ij}$ is the number of nearest neighbour ions, $N_A$ is the Avogadro number, g is the Lande g factor and $\mu_B$ is Bohr Magneton.

So, in-terms of Curie constant and field coefficients, $\chi_M$ can be written as

$$\chi_M = \frac{(C_{Fe_{HS}} + C_{Fe_{LS}} + C_G)T^2 + (2C_{Fe_{HS}}C_G n_{Fe_{HS}G} + 2C_{Fe_{LS}}C_G n_{Fe_{LS}G})T + C_{Fe_{HS}}C_{Fe_{LS}}C_G(2n_{Fe_{HS}G} n_{Fe_{LS}G} - n^2_{Fe_{HS}G} - n^2_{Fe_{LS}G})}{T^3 - (C_{Fe_{HS}}C_G n^2_{Fe_{HS}G} + C_{Fe_{LS}}C_G n^2_{Fe_{LS}G})T}$$

Since, $S_{Fe_{LS}} = 0$, $\chi_M$ reduces to $\chi_M = \frac{(C_{Fe_{HS}}+C_G)T^2 + (2C_{Fe_{HS}}C_G n_{Fe_{HS}G})T}{T^3 - (C_{Fe_{HS}}C_G n^2_{Fe_{HS}G})T} \quad \ldots\ldots(7)$

The exchange interaction constant $J_{ex,ij}$ based on spin Hamiltonian $H = -J_{ex,ij} S_i \cdot S_j$ is related to the transition temperature $T_C$ by the relation

$$T_c = \frac{\left(J_{ex,Fe_{HS}G}\right)[Z_{Fe_{HS}G} Z_{GFe_{HS}} S_{Fe_{HS}}\left(S_{Fe_{HS}}+1\right) \cdot S_G(S_G+1)]^{1/2}}{3k_B} \quad \ldots\ldots(8)$$

Here $S_i$ are the spin quantum number ($S_{Fe_{HS}} = 2, S_G = 1/2$), $Z_{ij}$ is the number of nearest neighbours which is 2 in this case. For HC, we obtained $T_c$=-36K, when used in this equation, gives a negative exchange constant of larger magnitude $\frac{J_{HC,Fe_{HS}G}}{k_B} = -25.4$. While, in-case of LC, using $T_c=0.82K$, gives a positive exchange constant of $\frac{J_{LC,Fe_{HS}G}}{k_B} = 0.58$. Hence, the negative exchange coupling constant of HC symbolizes antiferromagnetic interaction (anti-cooperative) between Fe HS and rGO magnetization state while positive exchange coupling constant for LC depicts ferromagnetic type (cooperative) interaction. Their relative magnitude signifies their strength of interactions.

**Transport measurement Methods:**

For transport measurement we prepared electrical contact pad on a Si/SiO$_2$ (300 nm) wafer. Ti/Au (10 nm/50 nm) metals were deposited using shadow mask in thermal evaporator to fabricate contacts without any resist residue. The separation between the electrodes is about 25μm. The rGO-SCO hybrid solutions were then drop casted to form a continuous layer between the electrodes. Temperature dependent electrical transport were carried out inside a dipstick in an inert environment (Ar 99.9%). The temperature of the sample was monitored using lakeshore 360 temperature controller. For electrical transport measurement, a constant ac voltage bias of 1 V at a frequency of 83.6 Hz was applied to the as prepared device, and the corresponding current was recorded using Lock-in technique.